\documentstyle[twocolumn,epsfig,aps]{revtex}

\begin{document}
\title{Investigation of the Mesoscopic Aharonov-Bohm effect\\
in Low Magnetic Fields}
\author{A.E. Hansen, S. Pedersen, A. Kristensen,\\
C.B. S\o rensen, and P.E. Lindelof\\
The Niels Bohr Institute, University of Copenhagen,\\
Universitetsparken 5, DK-2100 Copenhagen, Denmark}

\maketitle

\begin{abstract}
We have investigated the Aharonov-Bohm effect in mesoscopic semiconductor GaAs/GaAlAs rings in
low magnetic fields. The oscillatory magnetoconductance of these systems 
is systematically studied as a function of electron density. We observe phase-shifts of 
$\pi$ in the magnetoconductance oscillations, and halving of the fundamental $h/e$ 
period, as the density is varied. Theoretically we find agreement with the experiment, 
by introducing an asymmetry between the two arms of the ring.
\end{abstract}

PACS numbers:73.23-b, 73.20.Dx, 72.15.Rn

\section{Introduction}

The Aharonov-Bohm (AB) effect, first proposed in 1959 \cite{ab}, was experimentally realized in a normal metal system in 1982 
\cite{sharvin}.
Later the AB effect was observed in a semiconductor
system \cite{timp1}, and was the subject of a number of investigations
which expanded our general understanding of mesoscopic physics
\cite{timpagain,ford,ismail}. These investigations focused their
attention at relatively high magnetic fields. Only
\cite{chris} directly adressed the phase of the oscillations.

Recently, due to the perfection of e-beam lithography,  the AB effect has been the subject of new interest.  AB rings are now 
used to perform phase sensitive measurements  on e.g.\ quantum dots \cite{heiblum} or on rings were a local gate only affects the properties
in one of the 
arms of the ring \cite{mailly}. The technique used in these reports is to locally change the properties of one of the arms in the 
ring, and study the AB effect as a function of this perturbation. Information
about the changes in phase can be extracted from the measurements. Especially the observation of a period halving from $h/e$ to $h/2e$ 
and phase-shifts of $\pi$ in the magnetoconductance signal has attracted large interest.

\section{Experiment}

\begin{figure}
\centerline{
\epsfig{file=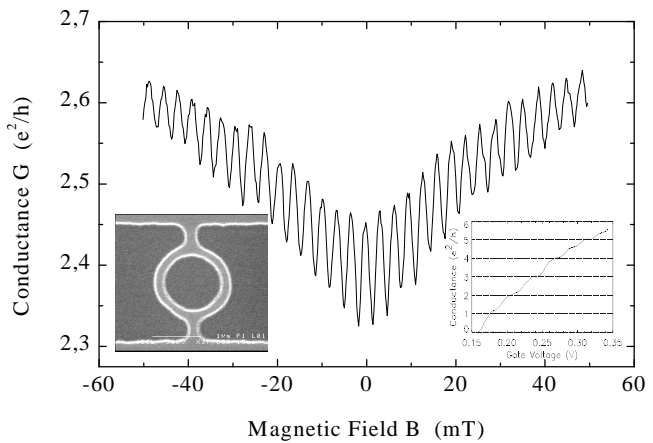,width=11cm}
}
\caption{
Measured magnetoconductance of the device shown on the SEM picture in the left insert. 
The magnetoconductance show very clear AB oscillations superposed on a slowly varying background. 
The right insert displays the zero 
magnetic field conductance at $T=4.2 {\rm K}$ as function of gate voltage. The
conductance curve displays distinct steps which show that the device is in a few-mode regime.
}
\label{1}
\end{figure}

We fabricate the AB rings in a standard two dimensional electron gas (2DEG) situated 90nm below the surface of a 
GaAs/GaAlAs heterostructure.  The 2DEG
electron density is $2.0 \cdot 10^{15}\rm{m}^{-2}$ and the mobility is $90\rm{T}^{-1}$. This corresponds to a mean free path of 
app.\ $6$ $\rm{\mu m}$. 
The ring is defined by e-beam lithography and realized with a shallow etch technique \cite{Anders}.
The etched AB structure has a ring radius of $0.65 \rm{\mu m}$ and a width of the arms of $200\rm{nm}$ 
(Fig. \ref{1}, left insert).
A $30\rm{\mu m}$ wide gold gate covers the entire ring, and is used to change the electron density.
A positive voltage $V_g$ must be applied to the gate for the structure to conduct.
The sample was cooled  to $0.3{\rm K}$ in a $^{3}$He cryostat equipped with a copper electromagnet. 
Measurements were performed using a conventional voltage biased lock-in technique
with a excitation voltage of $V=7.7{\rm \mu V}$ oscillating at a frequency of $131 {\rm Hz}$. 
Here we show measurements performed on one device, similar results 
have been obtained with another device in a total of six different cool-downs.

\section{Results}

We first consider the conductance as a function of the voltage applied to the global gate at zero magnetic field.
This is shown in Fig.\ \ref{1} (right insert), at $T$=$4.2 {\rm K}$. Steps are observed at approximate integer values of 
$e^{2}/h$. At least four steps are seen as the voltage is increased with $0.18{\rm V}$ from pinch-off. Such steps have previously been 
reported in AB rings \cite{ismail}. The steps show that our system, in the gate voltage regime used here, only has a 
few propagating modes. When the temperature is lowered a fluctuating signal is superposed the conductance curve. At $0.3$K,
the steps are completely washed out by the fluctuations. We ascribe the fluctuations to resonances. 
They appear at the temperatures where the AB
oscillations become visible and are the signature of a fully phase coherent device.

\begin{figure}
\centerline{
\epsfig{file=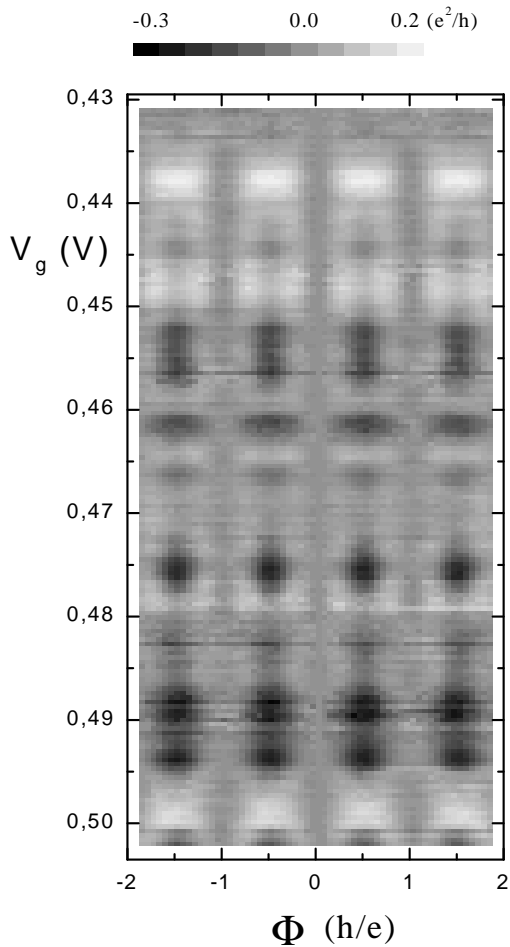,width=10cm}
}
\caption{Grayscale plot of the measured conductance subtracted the conductance at zero field,
$G(\Phi,V_{G})$-$G(0,V_{G})$, as a function of applied magnetic
flux $\Phi$ through the ring (horizontal axis) and global gate voltage $V_{g}$
(vertical axis).
}
\label{gray}
\end{figure}

We show in Fig.\ \ref{1} an example of a magnetoconductance measurement.
Here the amplitude of the oscillations is $\sim$ 7 \% around zero field. We have seen oscillation amplitudes of up to 10 \%.

The conductance measurement is, due to a long distance from the voltage probes to the sample, effectively two-terminal. 
Hence the magnetoconductance must be symmetric, $G(B)$ = $G(-B)$,
due to the Onsager relations. Here $B$ is the applied magnetic field. This means, that there can only be a maximum or
a minimum of the conductance at zero field, or stated differently, that the phase of the oscillations is $0$ or $\pi$ 
\cite{heiblum}.

\begin{figure}
\centerline{
\epsfig{file=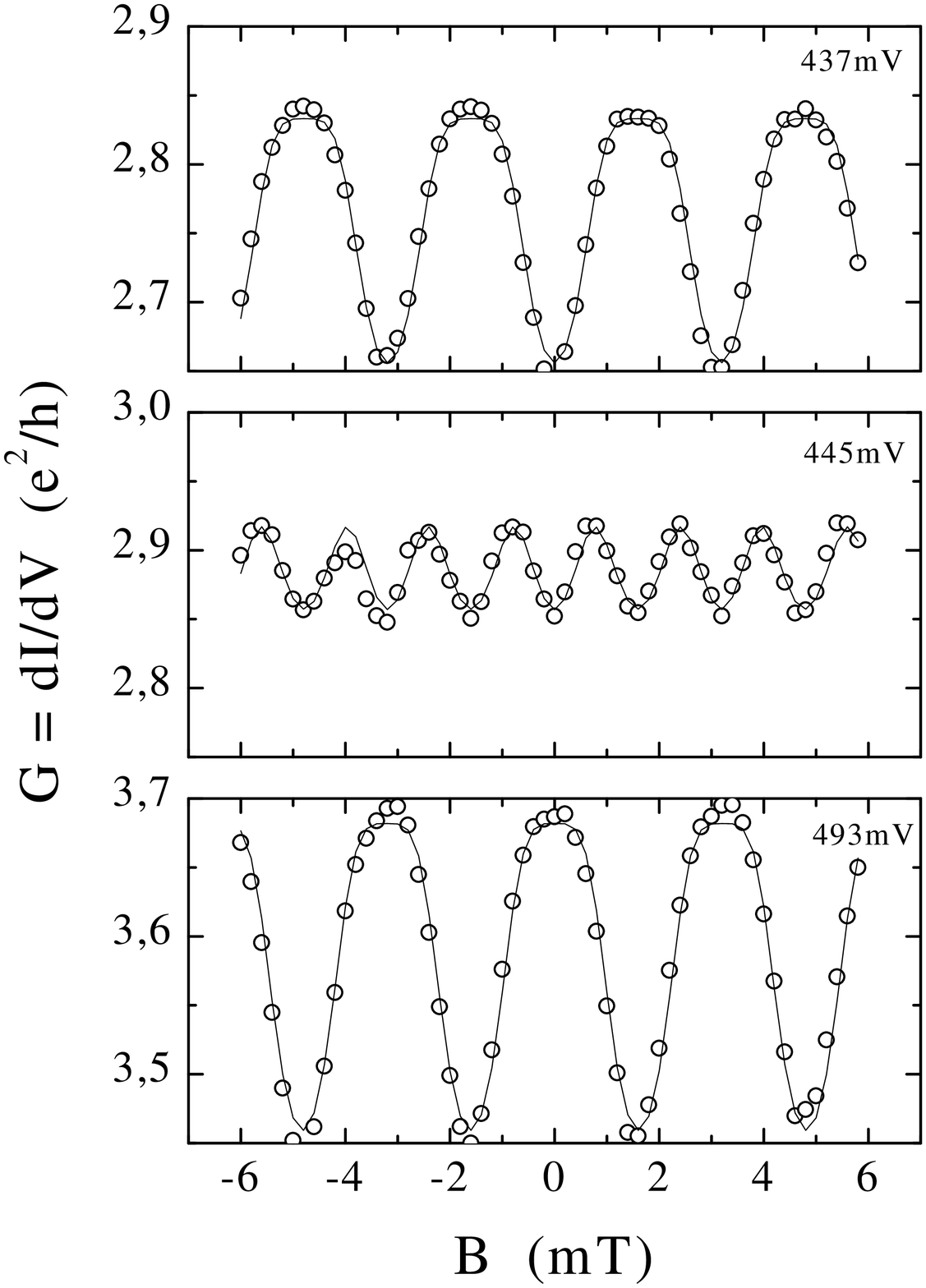,width=\linewidth}
}
\caption{Examples of magnetoconductance traces, showing (from above) AB phase of $\pi$,
period halving, and  AB phase of $0$. The voltages refer to the $V_g$-axis on the
previous Fig.\ \ref{gray}. Circles are measurements, lines are fits with Eq.\ \ref{fiteq}.
The values of $\delta$ obtained from the fit are, from above, $1.50$, $0.73$, and $0.00$.
}
\label{ex}
\end{figure}

In Fig.\ \ref{gray} we show the conductance $G(\Phi,V_g)$ {\sl subtracted} the fluctuating conductance at zero field $G(0,V_g)$,
as a function of magnetic flux $\Phi$ through the ring and gate voltage $V_g$.
The conductance is symmetric. Note that the dark (light) regions 
correspond to magnetoconductance traces with an AB phase of $0$ ($\pi$). To exemplify this,
we show single traces in Fig.\ \ref{ex}. Another remarkable feature is the occurrence of traces with {\sl half} the 
expected period in magnetic flux, see Fig.\ \ref{ex}.
We observe phase-shifts in the magnetoconductance, and occasional halving of the period, in all our measurements.
The transitions between situations with AB-phase $0$ and $\pi$ are smooth as the gate voltage is changed, 
as can be seen in Fig.\ \ref{gray}. In between, magnetoconductance traces
that have both $h/e$- and $h/2e$-periodicity appear, Fig.\ \ref{fits}.

The zero-field conductance $G(0,V_g)$ for the measurement shown in Fig.\ \ref{gray} varies between $2.5$ to $4.5$ in units
of $e^2/h$. We find in general, that for conductances of the AB ring less than app.\ $2e^2/h$, the AB oscillations are
weak or not present at all. This might be due to one of the arms pinching off before the other.

\section{Discussion}

The fact that the AB oscillations can have a minimum at zero field, implies that the AB ring on these occasions
is not symmetric, in the sense that the quantum phase acquired by traversing the two arms is not the same. In order 
to understand the behaviour, we compare our measurements with the theory \cite{bil1,bil2}, which is derived for a phase coherent
device with 1D independent electrons, and only one incident mode. This is the simplest possible theoretical model
one can think of. The conductance $G$ is given by \cite{bil2}
\begin{eqnarray}
\label{bil}
{\rm G}(\theta, \phi, \delta )=& & \frac{2e^{2}}{h} 2\epsilon {\rm g}(\theta,\phi) \nonumber\\
& &(\sin^{2}\phi\cos^{2}\theta+\sin^{2}\theta\sin^{2}\delta-\sin^{2}\phi\sin^{2}
\delta).
\end{eqnarray}
Here, $\phi=k_{F}L$, where $k_{F}$ is the Fermi wave number and $L$ is half the circumference
of the ring, is the average phase due to spatial propagation.
$\delta=\Delta (k_{f}L)$ is the phase difference between the two ways of traversing the ring.
When $\delta$ is not $0$, the AB oscillations might be phase-shifted by $\pi$.
$\theta=\pi \Phi/\Phi_{0}$ is the phase originating from the magnetic flux.
The coupling parameter $\epsilon$ can vary between $0$, for a closed ring, and $1/2$.
The function  ${\rm g(\theta, \phi)}$ is given by
\begin{eqnarray}
& &{\rm g(\theta, \phi)}= \nonumber\\
& &\frac{2 \epsilon}{(a_{-}^{2}\cos2\delta+a_{+}^{2}\cos2\theta-(1-\epsilon)\cos2\phi)^{2}+\epsilon^{2}\sin^{2}2\phi},
\end{eqnarray}
where $a_{\pm}=(1/2)(\sqrt{1-2\epsilon}\pm1)$.
Overall, we find the best agreement with the lineshape of the oscillations by taking $\epsilon$ = $1/2$, as
expected for an open system. Previously \cite{prb}, we estimated $\phi = k_FL$ $\sim$ 
$100-160$, for the gate voltage regime used here. However, note that varying $\phi$ and $\delta$ between $0$ and $\pi/2$ 
in the expression (\ref{bil}) exhaust all possible lineshapes of the magnetoconductance oscillations.

The equation (\ref{bil}) gives a conductance that can oscillate between $0$ and $2(e^2/h)$. The scale of the oscillations
as seen in in Fig.\ \ref{gray}, is at most $0.3(e^2/h)$. In order to match the lineshape of the magnetoconductance
oscillations to measurements, we use the form
\begin{equation}
\label{fiteq}
{\rm G(B)}={\rm G_o}+{\rm G_{\Delta}}\cdot  {\rm G}(\theta(B), \phi, \delta )_{\epsilon=1/2}.
\end{equation}
The introduction of the parameters ${\rm G_o}$ and ${\rm G_{\Delta}}$ is partly justified by the fact that 1) the experiment is performed
at a finite temperature, where the device might not be perfectly coherent. Incoherent transmission will on average not contribute
to the magnetoconductance oscillations. 2) For a system with more than one incident mode, again there will be a constant background and
the amplitude of the oscillations will be diminished \cite{bil2}.

\begin{figure}
\centerline{
\epsfig{file=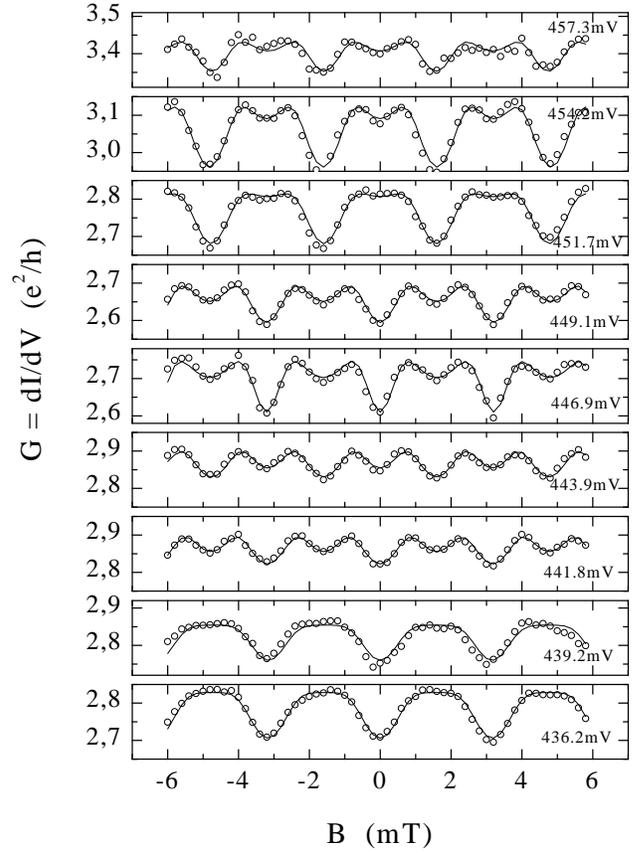,width=\linewidth}
}
\caption{Several magnetoconductance traces. Circles are measurements, lines are fits with Eq.\ \ref{fiteq}.
}
\label{fits}
\end{figure}

The lines in Fig.\ \ref{ex} are fits with the form (\ref{fiteq}). Note first, that indeed the expression (\ref{bil}) can
produce  both phase-shifts and halving of period. (For $\phi=\delta=\pi/4$ the period is purely $h/2e$.) Next, in Fig.\ \ref{fits}
several magnetoconductance traces are fitted with (\ref{fiteq}). The lineshapes of (\ref{bil}) agree nicely with the measurements.
Note however, that the introduction of the two extra parameters ${\rm G_o}$ and ${\rm G_{\Delta}}$ in (\ref{fiteq}), in addition
to $\phi$ and $\delta$ gives $4$ adjustable parameters in the fit. In order to extract solid information on the variation of
$\phi$ and $\delta$ in the experiment, an independent assesment of ${\rm G_o}$ and ${\rm G_{\Delta}}$ will be needed.

\section{Conclusion}

The oscillatory magnetoconductance of an AB ring, and in particular the phase of the oscillations, is systematically 
studied as a function of electron density. We observe phase-shifts 
of $\pi$ in the magnetoconductance oscillations, and halving of the fundamental $h/e$ period, as the density is varied.
All those features are reproduced by a simple theoretical model \cite{bil2}, when allowing for an asymmetry in the electron density in the 
two arms of the ring. 
Our interpretation gives a
simple explanation for why period-halving and phase shifts should appear in mesoscopic AB rings.
Further, our measurements suggest that variations in single-mode characteristics might be probed
by studying the lineshape of the AB oscillations.

\section{acknowledgements}
We wish to thank David H.\ Cobden and Per Hedeg\aa rd for enlightening discussions.
This work was financially supported by Velux Fonden, Ib Henriksen Foundation, Novo 
Nordisk Foundation, the Danish Research Council (grant 9502937, 9601677 and 9800243) 
and the Danish Technical Research Council (grant 9701490)


\begin{thebibliography}{99}
\bibitem{ab} Y. Aharonov and D. Bohm, Phys. Rev, {\bf 115} (1959) 485.
\bibitem{sharvin} D.Yu. Sharvin, and Yu.V.Sharvin, JETP  Lett. {\bf 34} (1982) 2
72.
\bibitem{timp1} G.\ Timp et al, Phys. Rev. Lett. {\bf 58} (1987) 2814
\bibitem{timpagain} G.\ Timp et al., Phys. Rev. B {\bf39} (1989) 3491
\bibitem{ford} C.J.B.\ Ford et al., J.Phys: Solid State Phys. {\bf 21} (1988) L325
\bibitem{ismail} K.\ Ismail et al., Appl. Phys Lett. {\bf 59} (1991) 1998; J.\ Liu et al., Phys. Rev B {\bf 47} (1993) 13039; J.\ Liu et al., Phys. Rev. B {\bf 48} (1993) 15148; J.\ Liu et al., Phys. Rev. B {\bf 50} (1994) 17383.
\bibitem{chris} C.J.B.\ Ford et al., Appl.\ Phys.\ Lett {\bf 54} (1989) 21; C.J.B.\ Ford et al., Surf.\ Sci {\bf 229} (1990) 307
\bibitem{heiblum} A.\ Yacoby et al., Phys. Rev. Lett. {\bf 73} (1994) 3149; A.\ Yacoby et al., Phys. Rev. Lett. {\bf 74} (1995) 4047; A.\ Yacoby et al., Phys. Rev. B. {\bf 53} (1996) 9583; R.\ Schuster et al., Nature {\bf 385} (1997) 417; E.\ Buks et al., Nature {\bf 391} (1998) 871.
\bibitem{mailly}  G.\ Cernicchiaro et al., Phys. Rev. Lett {\bf 79} (1997) 273
\bibitem{Anders} A.\ Kristensen et al., Physica B {\bf 249-251} (1998)180-184; A.\ Kristensen et al., J. Appl. Phys {\bf 83} (1998) 607.
\bibitem{bil1} M. B\"uttiker et al., Phys. Rev. B {\bf 30} (1984) 1982; Y. Gefen et al., Phys Rev. Lett. {\bf 52} (1984) 129.
\bibitem{bil2} M.\ B\"uttiker, SQUID'85 - Superconducting Quantum Interference Devices and their Applications, edited by H.D. Haklbohm and H. L\"ubbig (Walter de Gruyter, Berlin, New York 1985) page 529. 
\bibitem{prb} S. Pedersen et al., submitted to Phys. Rev. B, available from cond-mat/9905033.
\end{thebibliography}
\end{document}